\documentclass[manuscript]{aastex}

\newcommand{\kms}{km~s$^{-1}$}
\newcommand{\gal}{$\alpha$}
\newcommand{\gb}{$\beta$}

\newcommand{\etal}{et al.}
\newcommand{\smpy}{M$_\odot$~yr$^{-1}$}

\slugcomment {ApJ Letters, in press}

\shorttitle{Helium Omega Cen}
\shortauthors{Dupree and Avrett}

\begin{document}


\title{Direct Evaluation of the Helium Abundances in Omega Centauri}


\author{A. K. Dupree and E. H. Avrett}

\affil{Harvard-Smithsonian Center for Astrophysics, Cambridge, MA
  02138, USA}
\email{dupree@cfa.harvard.edu; eavrett@cfa.harvard.edu}


\begin{abstract}
A direct measure of the helium abundances from the near-infrared
transition of \ion{He}{1} at 1.08$\mu$m  is obtained for two nearly
identical red giant stars in the
globular cluster \object{Omega Centauri}. One star exhibits
the \ion{He}{1} line; the line is weak or absernt in the other
star.  Detailed non-LTE semi-empirical models
including expansion in spherical geometry are developed 
to match  the chromospheric H\gal, H\gb, and 
\ion{Ca}{2}~K lines, in order to predict the helium profile and derive
a helium abundance.  The red giant spectra suggest a 
helium abundance of 
$Y \le 0.22$ (LEID 54064) and $Y=0.39-0.44$ 
(LEID 54084) corresponding to  a difference
in the abundance $\Delta Y \ge 0.17$.
Helium is enhanced in the giant star (LEID
54084) that also contains enhanced aluminum and magnesium.  
This direct  evaluation of the helium abundances 
gives observational support to the theoretical conjecture that multiple
populations harbor enhanced helium in addition to light elements that 
are products of high-temperature hydrogen burning.  We demonstrate
that the 1.08$\mu$m \ion{He}{1} line can yield a
helium abundance in cool stars when constraints on the semi-empirical
chromospheric model are provided by other spectroscopic features.

\end{abstract}

\keywords{stars: individual (\object[Cl*NGC5139 LEID54064]{LEID
    54064}, \object[Cl*NGC5139 LEID54084]{LEID 54084} ) - stars:
  abundances - stars: atmospheres - globular clusters: individual
  (Omega Centauri)}

\section{Introduction}
Our current understanding of stellar populations in globular
clusters has dramatically changed with the discoveries of 
multiple stellar generations in a single globular cluster.
While variations in color and a spread in the [Fe/H] values
of red giants in massive clusters have been long recognized 
(Woolley 1966, Geyer 1967) along with variations  of
light elements (Martell 2011), the firm identification
of multiple populations on the main sequence in Omega Centauri  
(Anderson 1997; Bedin et al. 2004; Bellini et al. 2010), and subsequently
several other clusters (cf. Gratton et al. 2012), was surprising and 
continues to present 
theoretical challenges.  Norris (2004)  
suggested, based on  isochrone calculations, that dwarf stars
on the `blue' main sequence in Omega Cen would be enhanced 
in helium by $\Delta$Y$\sim$0.10$-$0.15. 
The lowered hydrogen opacity   causes stars of the same
mass to appear hotter and more luminous (Valcarce \etal\ 2012).  
Subsequently, the assessment of metals
in dwarfs on the bifurcated main sequence in Omega Cen,
showed that the hotter objects  (the `blue' dwarfs) were
less metal-poor than the  `red'  dwarf stars (Piotto
\etal\ 2005).  Stellar  models  suggest that
increased metals also signal the presence of  enhanced helium 
in the `blue' main sequence. The source (or sources)
of such an enhancement remains elusive. One attractive
explanation appears to be a second stellar generation formed from the
material lost by the first generation of intermediate mass stars
during their asymptotic giant phases (D'Ercole et al. 2010; Johnson \&
Pilachowski 2010; Renzini 2013), although
other possibilities such as fast-rotating massive stars (Charbonnel
\etal\ 2013) or massive binary-star mass overflow (de Mink \etal\
2009) may well contribute  
(cf. Gratton \etal\ 2012).  The
formation of  cluster populations with  several generations
of star formation also impacts an understanding of the
halo of the Milky Way, satellites of our Galaxy, and
the star formation and assembly history of other galaxies 
(cf. Gratton \etal\ 2012; Brodie \& Strader 2006).

It is obviously of great interest if a helium enhancement
could be verified in globular cluster stars in our Galaxy.  
A direct measure of the helium abundance from a spectrum would
provide confirmation of Norris' conjecture. Such a measurement is challenging
because useful lines of  helium are generally absent in 
the optical spectra of cool stars.  Moreover, in hotter
stars, such as blue horizontal branch objects, sedimentation caused by
diffusion and element
stratification occur.   Helium abundances from the spectroscopy
of hot horizontal branch stars in Omega Cen demonstrate the effects
of surface diffusion, or mixing during late helium core
flashes (DaCosta \etal\ 1986; Moehler 
\etal\ 2011; Moni Bidin \etal\ 2012)   and derived 
abundance values vary widely from Y $\le$0.02 to Y=0.9.
 
In cool stars, a transition in \ion{He}{1} occurs in the near-infrared 
at 1.08$\mu$m  and  has been identified in many 
metal-poor field stars, where, in addition to abundances,
it can indicate atmospheric dynamics because the lower level
of the transition is metastable (Dupree \etal\ 1992, 2009; Smith \etal\ 2012). 
In Omega Centauri, a closely matched group of first-ascent red giant stars displays
strong and weak helium absorption that correlates (Dupree \etal\ 2011)
with increased [Al/Fe] and [Na/Fe] abundance, more than with [Fe/H]. This
result  gave direct observational support to the idea that products
of high-temperature hydrogen burning in a previous stellar 
generation had, in fact, occurred.  A quantitative
measure of the helium abundance in these objects 
is the goal of this Letter.   

Pasquini \etal\ (2011) calculated profiles of   
the \ion{He}{1} 1.08$\mu$m line in an approximate way  
based on a stationary plane-parallel  model 
applied to two  very cool luminous stars in NGC 2808.  
They showed that a change in the chromospheric structure itself
can strengthen or weaken helium absorption. In fact,
chromospheric line profiles are highly sensitive to the structure and
dynamics of the atmospheric model.  In this paper, we
have  selected similar stars and  
first constrained the atmospheric structure and dynamics
using other chromospheric lines.  A  model for the
radiative transfer must be used that is appropriate to the stars.
Following that,  the abundance
of helium can be inferred from line synthesis using the 
semi-empirical atmospheric model that is anchored by other
chromospheric lines. 

Here we focus on two `identical' red giants in Omega
Centauri, LEID 54064 and LEID 54084 (van Leeuwen \etal\ 2000).
They are located $\sim$5.7 arc min to the SW from the cluster center,
and are separated by 1.6 arcminutes on the sky. 
These giants have  very similar  temperatures, luminosities,
and values of [Fe/H] (Table 1).  However they differ remarkably in 
[Na/Fe] and [Al/Fe] abundances and the strength of the helium line
(Dupree \etal\ 2011).  The star LEID 54084 exhibits enhanced light
elements as compared to LEID 54064. 

\section{Modeling Chromospheric Lines}

The PANDORA code (Avrett \& Loeser 2003, 2008) is used to develop
the semi-empirical, spherical model of the chromosphere where the temperature
distribution, the turbulent velocities, and the
expansion velocities are adjusted to obtain optimum agreement
between calculated profiles and observations of chromospheric lines (H$\alpha$,
H\gb, and \ion{Ca}{2}-K).  The initial model 
consists of a static LTE photosphere corresponding to a effective
temperature of 4740K (Kurucz 2011), gravity $log~g$= 1.75, a 
 stellar radius of 20$R_\odot$, 
and [Fe/H]=$-$1.72 with the $\alpha$-abundances enhanced by
$+$0.44 dex. Chromospheric line emission is 
essentially unaffected by the photospheric model. 
A chromospheric structure similar to  
other metal-poor models (M\'esz\'aros \etal\ 2009) was
added to begin the iterations, and expansion started in
the low chromosphere.  Our calculations
assume multi-level atoms  (\ion{H}{1}:15 levels, \ion{Ca}{2}: 5
levels,  
\ion{He}{1}: 13 levels), and the iterations explicitly 
consider the velocity field
in the evaluation of the line source functions and as a contribution
to the pressure in the hydrostatic equilibrium equations.  The total model is 
iterated  with full and complete non-LTE calculations in order to
match the chromospheric line profiles.  The \ion{Ca}{2}-K line profile
is computed with partial frequency redistribution; complete
frequency redistribution is used for the hydrogen 
and helium lines.  These flux profiles
are calculated with an integration over the apparent spherical stellar
disk including the extended chromosphere.

The profiles of the optical lines, H\gal, H\gb, and \ion{Ca}{2}-K
were taken from spectra obtained with the MIKE double echelle spectrograph 
(Bernstein \etal\ 2003) mounted on the Magellan/CLAY
telescope at Las Campanas Observatory.   These spectra were 
used previously to derive elemental abundances (Dupree \etal\ 2011).  
The spectra and the calculated stellar profiles  for H\gal, H\gb, 
and \ion{Ca}{2}-K are shown in Fig. 1.  The observed 
profiles are effectively identical between
the two giants, signaling that the activity levels of the stars
are similar.  The spectra are well matched by the calculated profiles. 
Note the asymmetry in the H\gal\ line core; the core is formed
higher in the atmosphere than the rest of the profile and
is sensitive to the outflow.  However the line itself is
narrow, and demands a relatively low turbulent velocity, which
increases with height in the chromosphere. 
The final model (Fig. 2) has a temperature that extends to 10$^5$K
(although such high temperatures do not affect the profiles
evaluated here), and  an outflow velocity that reaches 100 \kms,
which yields a mass outflow rate of $\sim 3\times 10^{-9}$ \smpy. 
This rate follows straightforwardly from the 
atmospheric model (Fig. 2)  and is proportional to $N v r^2$ in the chromosphere, 
where $r$ is the radial distance at which the wind has a velocity $v$, and $N$ is the 
hydrogen density. This value exceeds by a factor of 1.3--1.5 the rate estimated
from an extension of the M\'esz\'aros  \etal\ (2009) fit 
to H\gal\ profiles of cooler stars shown in their Fig. 10.  
For more luminous stars in the more metal-rich NGC 2808, Mauas \etal\ 
(2006) find values of $0.7 - 3.8\times10^{-9}$ \smpy\ from the H\gal\ line.  Field 
metal-poor giants, comparable in $M_{bol}$ to our targets possess a mass loss rate 
spanning $1.3 \times 10^{-9} - 10^{-8}$ \smpy (Dupree \etal\ 2009).   While mass loss rates 
have been measured (M\'eszaros \etal\ 2009) to vary with time by factors of 1.5 to 6 in metal-poor red giants with 
luminosity $log L/L_\odot \sim 3.0$, the values inferred from semi-empirical model fits 
are less by an order of magnitude than the Reimers (1977), Origlia \etal\ (2007), 
or the Schr\"oder \& Cuntz (2005) approximations.

This temperature and velocity  model (Fig. 2) is used to evaluate the profile
of the \ion{He}{1} 1.08$\mu$m line.  The near-ir \ion{He}{1} lines
measured with  PHOENIX on Gemini-S were reported earlier (Dupree \etal\ 2011).   
The populations, ionization fraction, and continuum emission, are
evaluated in separate models for each value of the
helium abundance, and the profile is calculated assuming spherical 
geometry in an expanding atmosphere. The contribution of the extended chromosphere 
can be noted in the weak emission present on the long 
wavelength side of the line.  The helium absorption extends
substantially towards shorter wavelengths due to 
scattering in the expanding atmosphere and is enhanced by 
the metastable nature of the lower level of the transition. 
The helium  lines are essentially P Cygni profiles since the
red giants have extended atmospheres. 
The population of the lower level of the 1.08$\mu$m 
transition peaks at T=18,000K, but
lies within a factor of two of its maximum value between
14,000 and 25,000K;  the
outflow velocity doubles over this temperature span.

Various values of the  helium abundance, from Y=0.15 to Y=0.50
[log(n$_{He}$/n$_H$) ranging from 10.65 to 11.4], were 
assumed and 9 models calculated. The abundance selected 
minimizes the residuals 
between the observed and calculated profiles.
The star LEID 54084 clearly
exhibits a broad helium line  which could  extend
to shorter wavelengths beyond the \ion{Si}{1} 
absorption at 1.027$\mu$m but is compromised by the presence
of the water vapor blend with \ion{Si}{1}. 
A value of Y=0.39 to Y=0.44 well represents the depth of the
observed profile representing the minimum range in the
residuals.   Helium is  not clearly detected in LEID 54064. The 
calculated profiles 
for Y$\le$0.22  give a minimum in the residuals, and we adopt this
value as an upper limit to Y.
Inspection of the helium profiles shows that
a value of Y=0.25   overpredicts the strength of the
line in LEID 54064, and  the residuals of the fit are larger than 
for Y=0.22.  
These simulations suggest that the helium abundance difference is 
$\Delta Y \ge 0.17$ between the
two stars.

\section{Discussion and Conclusions}
This spectroscopic value of helium from LEID 54084, namely
Y=0.39--0.44 can be compared to values obtained from models of stellar
structure and evolution. In Omega Cen, Norris (2004) estimated 
the presence of helium  from isochrones matching the 
lower main sequence with values of Y ranging from 0.23 to 0.38.    
Piotto \etal\ (2005) noted the blue main sequence could only
be matched with stellar
models with helium abundance ranging from  0.35 $<$ Y $<$ 0.45 and concluded that Y=0.38
best fit the ridgelines in the color-magnitude diagram of Omega Cen.
HST photometry of an outer field in the cluster (King \etal\ 2012),  
reveals a helium abundance for the blue main sequence of
Y=0.39$\pm$0.02.  Recent Yonsei-Yale isochrones for several subpopulations in Omega Cen
(Joo \& Lee 2013) suggest a range in Y from 0.38 to 0.41. 
Thus the spectroscopic value of helium for LEID 54084, a  star with
enhanced light element abundances 
is  in harmony with the abundance inferred from
stellar structure models. In a more metal rich cluster, NGC 2808, 
the approximate model of Pasquini et al 
(2011) suggested one star may have a similar value of Y=0.39 to 0.5.

The Y value for LEID 54064 where the helium line is
weak (or not detected) has an upper limit (Y $\le$ 0.22) that is slightly 
less than the cosmic value (Y=0.24).   These abundances
suggest the helium enhancement, $\Delta$Y, is $\ge$0.17. 
King \etal\ (2012)  concluded from plausible fits to the
color magnitude diagram of Omega Cen that $\Delta$Y$\sim$0.15 where
a value for the primeval abundance of helium (Y=0.24) was 
chosen  for the red main sequence.
Piotto \etal\ (2005) required 
$\Delta$Y=0.14 to explain the differences in metal abundances found for the
blue and red main sequences.
It is interesting to note that the Sun requires a 
helium abundance of Y=0.27$-$0.28 to match the solar luminosity,
but due to diffusion and settling, the helium abundance
in the envelope is less, Y=0.24$-$0.25 (Christensen-Dalsgaard 2002; 
Guzik \& Cox 1993), and Y=0.16  (corresponding to n$_{He}$/n$_H$=0.05)
in the steady-state solar wind (Kasper \etal\ 2007).
It may be that
spectroscopy will yield different values for the helium abundance from those inferred
from stellar isochrone models, although  currently we do not know  
if the characteristics of the solar abundance pattern occur in these
metal-poor giant stars.

The optical and near-infrared spectra used here were acquired about 
3 months apart, and a variation in the line profiles might occur.
However, these giants  have log $L/L_\odot~\sim$ 2.2, and
$M_V \sim -$0.45  and  lie on the red giant branch below 
the stars that exhibit H\gal\ wing emission. It is this emission  
which can vary in strength in first-ascent red giants (M\'esz\'aros \etal\ 2008;
Cacciari \etal\ 2004). The remarkable similarity of  H\gal, H\gb,  and \ion{Ca}{2}-K
profiles between the two giants suggests that  activity 
does not cause significant changes.  

Another consideration might be the presence
of X-rays or EUV emission from a high temperature plasma.  Because
neutral helium can be photoionized and then recombine preferentially
into the lower level of the 1.08$\mu$m line, this process would enhance
the strength of the observed helium line.  Red giants need substantial 
magnetic confinement of material to produce hot
plasma;  magnetic signatures in 
the spectra of similar stars have
not been detected, and the coronae appear absent (Rosner \etal\ 1995).   The slightly metal-poor 
K giant, \gal\ Boo, has a `tentative detection' (Ayres \etal\ 2003)
of X-rays but, if indeed  present, 
they are a factor of 10$^4$ weaker in $L_X/L_{bol}$
than the average solar value and would seem to have little effect on
the profile.\footnote{{\it CHANDRA} 
images of Omega Cen (Cool \etal\ 2013) do not reach   
faint sources ($L_X \lesssim 10^{29}\ erg\ s^{-1}$).  The 
identified optical counterparts
of the X-ray sources are  binaries,  and not the single red giants that
are targeted here.}    In  \gal\ Boo, the equivalent width of the 1.083$\mu$m line 
varies in absorption strength  which could be caused by wind variation
as well as chromospheric excitation conditions 
(O'Brien \& Lambert 1986).    Single  metal-poor red  giants in the field 
also display a very weak 1.08$\mu$m  absorption line, and though these
stars are optically brighter, they have not been detected in X-rays.  
Population I giants, which
generally exhibit X-rays, have stronger helium
absorption  as compared to  their metal-poor field counterparts (cf. Dupree
\etal\ 2009).  This suggests  the line is not influenced by X-rays 
in the metal-poor stars.  
The \ion{Ca}{2}-K lines are very similar in the two
giants (Fig. 1) indicating that these stars have similar chromospheres 
such that  X-rays would not be present in only one star causing the
strengthening of the helium  absorption.  Thus it does not appear likely that X-rays 
contribute to the line formation for the targets considered here.  Several epochs of
measurement would clearly be useful to determine if variation occurs in the helium lines.

Pasquini \etal\ (2011) carried out a similar calculation for two stars
in the globular cluster NGC 2808. 
The 2 luminous stars ($log L/L_\odot\ \sim 3.2$)  selected by Pasquini \etal\ (2011) have 
different levels of activity as indicated by the Ca~K line which 
underscores the ubiquitous variability  of such luminous giants. 
These differences demand different semi-empirical models for the two stars yet 
only one model was used; in addition the observations of the optical and infrared  
spectra were separated by some weeks which brings uncertainty when
modeling such luminous  active objects.  Computation of the line profiles in Pasquini et al. invokes 
models that do not adequately represent the stars nor the conditions in their 
atmospheres. The use of a plane-parallel  approximation is questionable when 
modeling a star of radius $\sim$84 R$_{\odot}$. The computation assumed a static 
atmosphere, and the authors  simply shifted the calculated line in wavelength
to match observations.  However, the spectra show that the chromosphere, as measured by H\gal,  
Ca K, and the helium line, exhibits signatures of outflow.  
We have taken our  model  
and calculated the helium profile under the same assumptions 
adopted by Pasquini \etal\ (2011)  (plane parallel and static)  for comparison 
to a model with the appropriate assumptions for these stars, 
namely spherical geometry and expanding.  The results of this 
calculation show substantial differences.    Not only does the 
spherical model exhibit emission, but the absorption is 
larger than the static model due to the expanding atmosphere. A larger
star, with extended chromosphere and/or wind,  might be expected to exhibit more
substantial changes.  For the same value of the helium abundance,
the equivalent width of the absorption 
in the expanding spherical model is larger by 5 to 19\% 
than the  static plane-parallel model depending on the value of
Y. (Here we assumed Y=0.28 and Y=0.44.) Thus, interpreting the observed  
profile formed in an expanding large 
giant star,  by 'matching it' to a static, plane parallel profile, 
as did Pasquini \etal\ (2011), will lead to an overestimate of the abundance of helium.  
Pasquini \etal\ (2011)  do not compare the computed profiles of 
Ca~K and H\gal, to the stellar spectra so the adequacy of the models 
is unknown. Consideration of all of
these facts indicates that the determination of the helium abundance
in Pasquini \etal\ (2011) must be approached with caution.

The targets selected in this paper are of much lower luminosity where variability 
is absent or greatly minimized. Moreover, the  2 stars are effectively identical 
in temperature, luminosity, iron abundance, activity, and in chromospheric 
features - with the exception of helium and enhanced Al and Mg. The treatment of 
the radiative transfer is state of the art with a spherical atmosphere, assuming 
an outflow, where the outflow is incorporated into the source function for the 
lines.

The abundance of helium and its variation between these two giant
stars in Omega Cen gives quantitative observational confirmation of a
helium enhancement to accompany the enhanced light metals.  The
near-ir line of \ion{He}{1} can provide a probe of the helium
abundance in cool stars when  
additional chromospheric profiles
are available to constrain the atmospheric structure and dynamics 
and appropriate radiative transfer calculations are employed.   

\acknowledgments
We are grateful to Bob Kurucz who calculated specific photospheric
models to initiate the calculations.

{\it Facilities:} 
\facility{Gemini:South (PHOENIX), Magellan:Clay (MIKE)}

\begin{figure}
\begin{center}

\vspace*{-0.3in}

\includegraphics[angle=0.,scale=0.7]{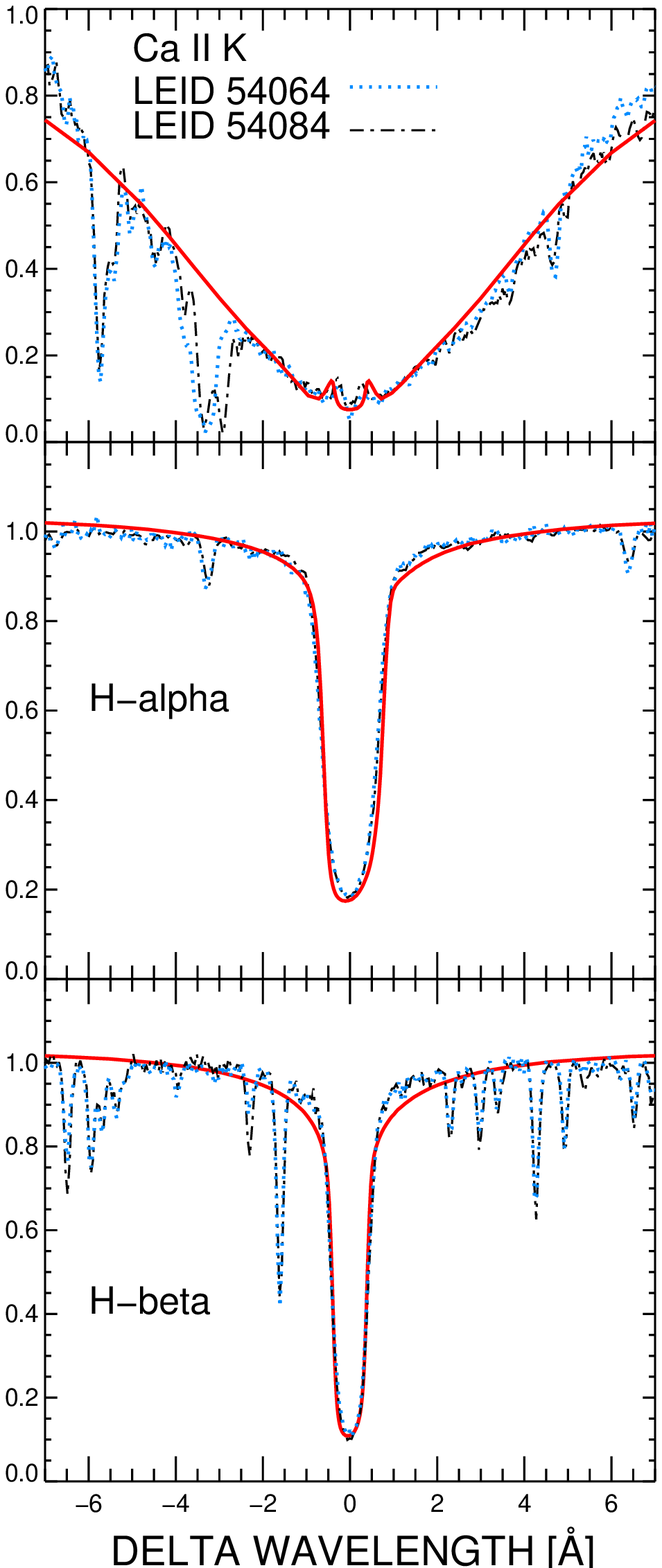}

\caption{H\gal\ , H\gb,  and \ion{Ca}{2} K-line region
   shown for the two giant stars with the model fit overlaid ({\it
   indicated by a solid line which is colored red in the electronic edition}). The spectra of the 
two stars are virtually identical. The
  \ion{Ca}{2} spectrum displays an interstellar absorption feature  
blended with the \ion{Fe}{1} line at
  $-$3\AA\ from the line core in the LEID 54064 spectrum, but distinct
from the \ion{Fe}{1} feature in the LEID 54084 spectrum. Weak emission
   may be present on the long-wavelength wing of the H\gal\ line and
   on both wings of the H\gb\ line. The success of the model can be
seen from the agreement between observed and calculated chromospheric
   profiles.  A color version of this figure is in the electronic edition.}
\end{center}
\end{figure}

\begin{figure}
\begin{center}
\includegraphics[angle=0,scale =0.7]{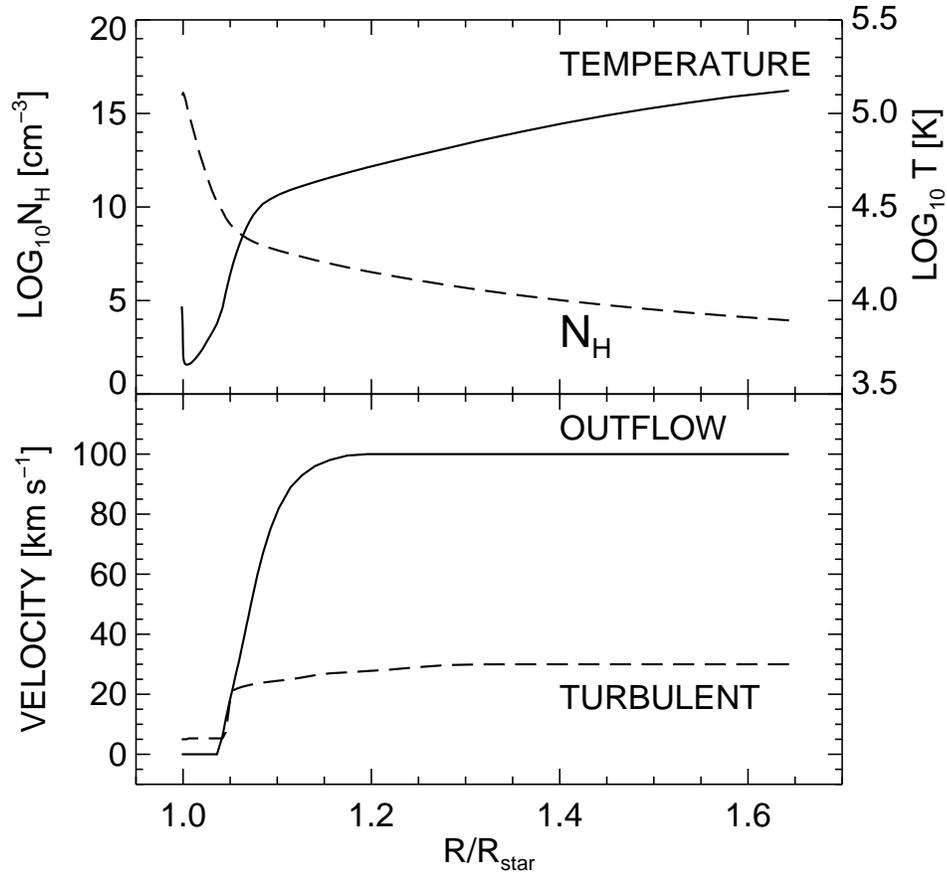}

\caption{Final model. The top panel displays the total hydrogen
density ({\it left axis}) and the temperature ({\it right axis}).
The lower panel shows the turbulent and outflow velocities needed
to match the observed profiles.}
\end{center}
\end{figure}

\begin{figure}
\begin{center}
\includegraphics[angle=90.,scale=0.5]{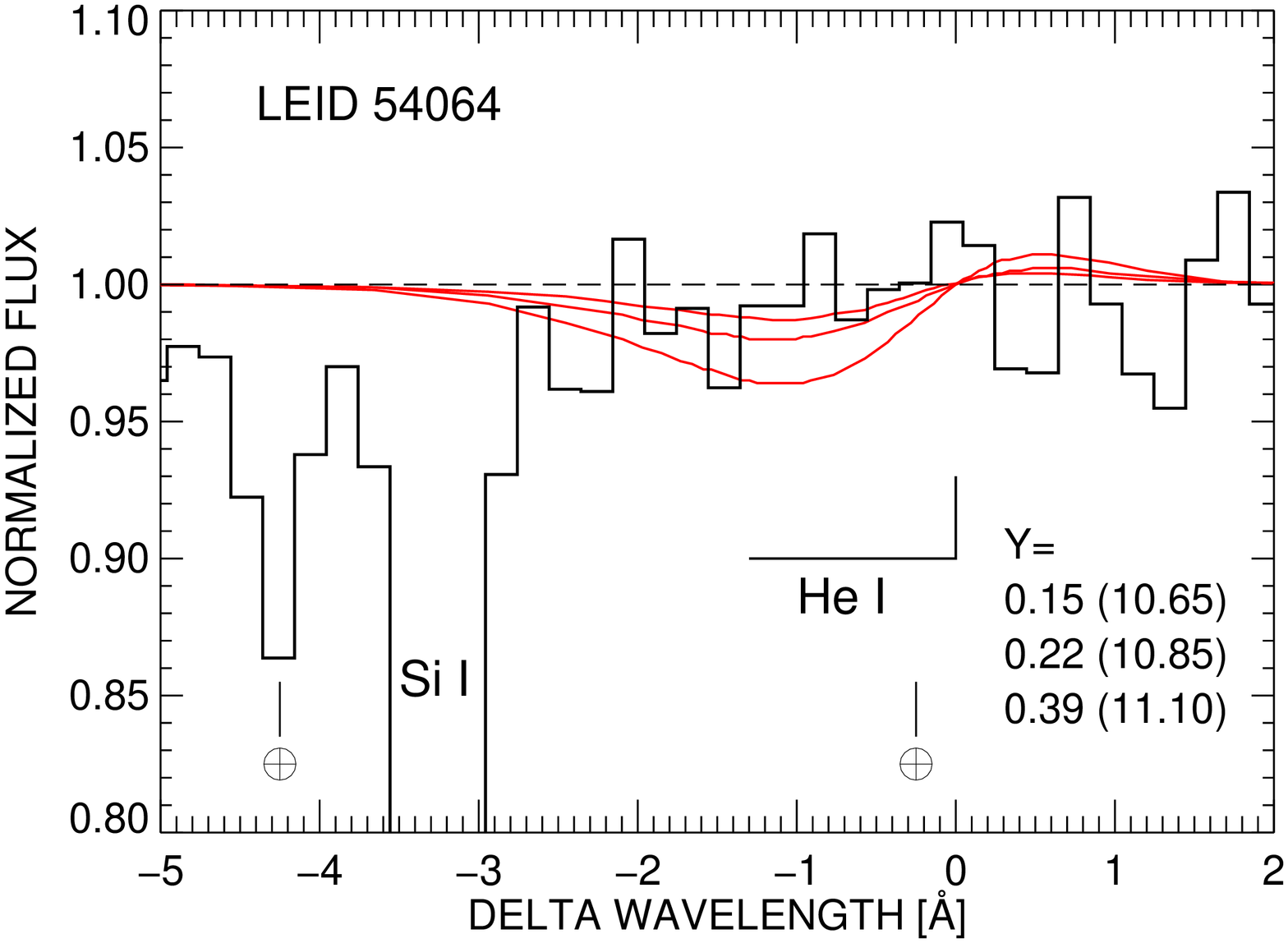}
\includegraphics[angle=90.,scale=0.5]{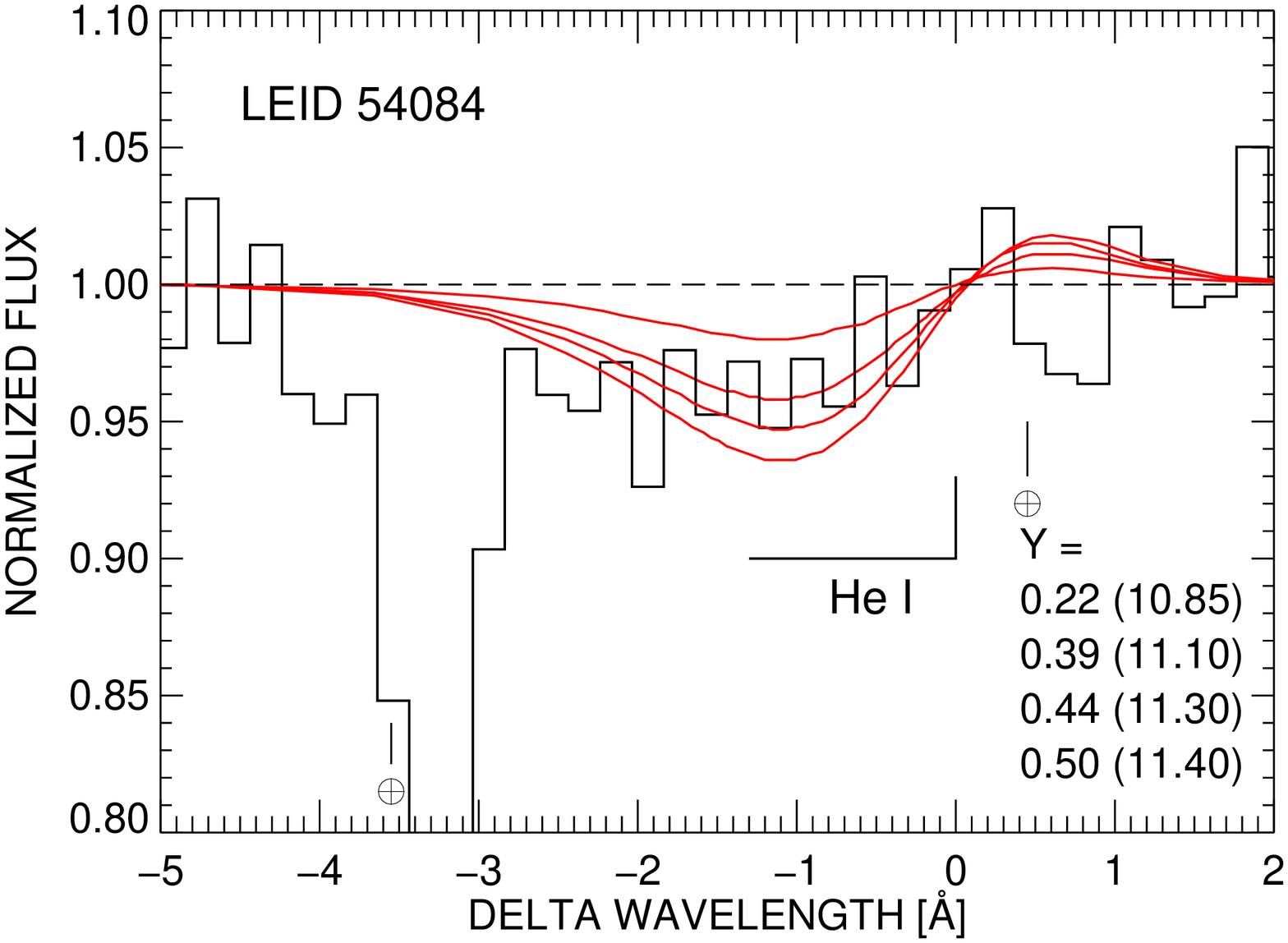}
\caption{Observed Helium lines binned to a resolution element 
in the two matched red giants with the
10830\AA\ profile as calculated for several values of the helium
abundance ({\it shown by a solid smoothly varying line that is colored red in the electronic edition}). 
Values of Y are given for the calculated curves arranged
top to bottom and the corresponding log
(n$_{He}$/n$_H$) is shown in parentheses where log $n_H$=12.00. }
\end{center}
\end{figure}


\begin{deluxetable}{lllc}
\tablecaption{Characteristics of Target Stars\label{tbl.stars}}
\tablewidth{0pt}
\tablecolumns{4}
\tablehead{
\colhead{Quantity}&\colhead{LEID 54064}&\colhead{LEID 54084}&\colhead{Refs.} \\
}
\startdata
V                   & 13.27 & 13.21 &  1 \\
B$-$V               & 1.048 & 1.044 &  1 \\
K$_s$               & 10.62  &10.56 &  2 \\
T$_{eff}$ [K]       & 4741  & 4745  &  3   \\
log g [cm s$^{-2}$] &1.76  & 1.74   &  3 \\
M$_V$               &$-$0.43 & $-$0.49 &4\\
log $L/L_\odot$     &2.21 & 2.23 & 5 \\
$[Fe/H]$              & $-$1.86& $-$1.79 &  3\\
$[Na/Fe]$             & $-$0.14& 0.37&  3\\
$[Al/Fe]$             & $\le$0.36 & 1.12 &  3\\
EW (\ion{He}{1}) [m\AA] & $\le$9.2& 89.5& 3\\
\enddata
\tablerefs {(1) van Leeuwen \etal\ 2000 ;(2) 2MASS All Sky Survey;
  Skrutskie \etal\ 2006; (3) Dupree \etal\ 2011; (4) Distance modulus 
from Johnson \&\ Pilchowski 2010; (5) Bolometric
correction from Alonso \etal\ 1999.  }

\end{deluxetable}

\end{document}